\documentclass[aps,prl,twocolumn,groupedaddress,showpacs]{revtex4}
\usepackage{graphics,graphicx}
\usepackage{color}
\usepackage{wrapfig}
\usepackage{amssymb,amsmath}
\usepackage{bookmath}

\newcommand{\Gts}{\Gamma_{{\rm s}}}

\newcommand{\req}[1]{Eq.~(\ref{#1})}
\newcommand{\reqs}[1]{Eqs.~(\ref{#1})}

\begin{document}
%~~~~~~~~~~~~~~~~~~~~~~~~~~~~~~~~~~~~~~~~~~~~~~~~~
\title{Spin relaxation in quantum dots due to electron exchange with leads}
\author{A.~B.~Vorontsov}
\author{M.~G.~Vavilov}
%\email[contact:]{anton@physics.wisc.edu}
\affiliation{Department of Physics,
             University of Wisconsin, Madison, Wisconsin, 53706, USA}
\date{October 21, 2008}
\pacs{73.23.-b, 75.30.Hx}

\begin{abstract}
We calculate spin relaxation rates in lateral quantum dot
systems due to electron exchange between dots and leads.
Using rate equations, we develop a theoretical description
of the experimentally observed electric current in the spin
blockade regime of double quantum dots.
Single expression fits the entire current profile
and describes the structure of both the conduction peaks
and of the suppressed (`valley') region.
Extrinsic rates calculated here have to be taken into account for
accurate extraction of intrinsic relaxation rates
due to the spin-orbit and hyperfine spin scattering mechanisms
from spin blockade measurements.
\end{abstract}
%~~~~~~~~~~~~~~~~~~~~~~~~~~~~~~~~~~~~~~~~~~~~~~~~~~~~~~~~~~~~~~~~~~~~~~~~~~~~~
\maketitle
%~~~~~~~~~~~~~~~~~~~~~~~~~~~~~~~~~~~~~~~~~~~~~~~~~~~~~~~~~~~~~~~~~~~~~~~~~~~~~

%\paragraph{Introduction.}
During the last decade, considerable progress was made in
development of devices utilizing spin degrees of freedom of
electron systems,\cite{hanson07}
promising manufacturing of functional devices.
The ultimate usage of spin devices for quantum information
processing\cite{loss98} would capitalize on potentially weak
coupling of spin systems with their environments.
This coupling can be of two origins: (i) intrinsic to material; and
(ii) device specific (geometric or extrinsic), due to interactions with leads,
substrate, etc.
Knowledge of spin relaxation and decoherence
resulting from the coupling to environment is crucial for evaluating
the limitations of the spin-based devices.

The intrinsic mechanisms responsible for spin relaxation
in semiconductor devices are the hyperfine interaction
with nuclear spins %\cite{erlingsson01,khaetskii02,merkulov02,desousa03}
and spin-orbit interaction, %\cite{khaetskii00,golovach04,tahan05}.
for a review see Refs.~\onlinecite{zutic04,hanson07}.
Relaxation rates were experimentally analyzed for quantum dot devices
on the basis of GaAs heterostructures. %\cite{hanson03,johnson05a}
The resulting coherence and spin flip times can be as short as
$10-100$ns in zero magnetic field, but are much longer than
microseconds in the field\cite{hanson07}. For silicon dot structures
such measurements have yet to  be done; however the two intrinsic
mechanisms are small in bulk Si, and these times are expected to be
orders of magnitude longer, $\gtrsim 1$second.\cite{tahan02,prada08}

%%%%%%%%%%%%%%%%%%%%%%%%%%%%%%%%%%%%%%%%%%%%%%%%%%%%%%%%%%%%%%%%%
\begin{figure}[t]
\includegraphics[width=0.85\linewidth]{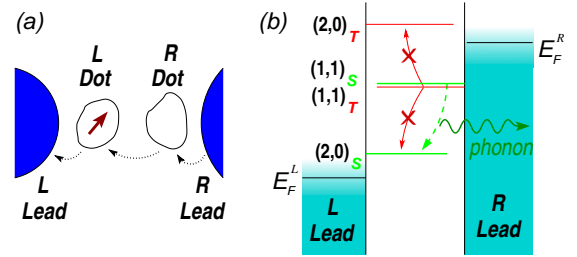}
\caption{ \label{fig:model}
    (Color online) (a) Lateral double dot system in the spin blockaded regime.
    In the initial configuration, the system has one electron on the left dot, $(1,0)$.
    Electric current through the dot flows from right to left in a sequence of steps.
    (b) Energy level representation of transport through the dot.
    The sequence of transitions via singlet states,
    $(1,0) \to (1,1)_{\rm S} \leadsto (2,0)_{\rm S} \to (1,0)$,
    quickly transports electron from the right to left lead,
    whereas current through intermediate triplet state,
    $(1,0) \to (1,1)_{\rm T}$, is blocked since subsequent transition
    to $(2,0)_{\rm S(T)}$ state is forbidden by spin (energy) conservation.
}
\end{figure}
%\noindent
%\hrulefill
%\end{wrapfigure}
%%%%%%%%%%%%%%%%%%%%%%%%%%%%%%%%%%%%%%%%%%%%%%%%%%%%%%%%%%%%%%%%%%%%%%%%

The goal of this paper is to analyze recent
studies~\cite{ono02,johnson05,shaji07,liu08} of charge transport
through double quantum dot systems. Such systems are proposed for
spin spectroscopy and control of dot states. The spin blockade,
Fig.~\ref{fig:model}, of electric current through double quantum
dots is one of the most striking signatures of the electronic spin
degree of freedom in nanoelectronic devices. We examine in detail
the electric current through a double quantum dot system. We point
out that while recent studies of relaxation processes in double dots
were focused on intrinsic mechanisms of spin relaxation,
the current %in the spin blockade regime
contains an essential contribution arising from the extrinsic
mechanism of spin relaxation as a consequence of electron tunneling
between dots and leads.
Below we propose a theoretical description of
this effect. %taking into account the spin relaxation through the leads.

We show that the current remains finite, and equal to $I=e\Gts/3$, even in the
absence of intrinsic spin relaxation mechanisms. Parameter $\Gts$
is the \emph{extrinsic} spin relaxation rate due
to electron tunneling to the leads.  This rate strongly depends on
the relative energy between the spin state in the dot, $E =
E_{\uparrow,\downarrow}$, and the Fermi level, $E_F$, of the
connected lead \bea \Gts = \Gamma_\sm{R} [1-f(\Delta)] + 2
\gamma_2(\Delta) \,, \label{eq:Tts}
%\ee
\\
%\be
\Delta =E-E_F \,,\quad
\gamma_2 = \Gamma_\sm{R}^2 \frac{T}{\Delta^2} \frac{\hbar}{2\pi}\,,
\nonumber
\eea
where $f(\Delta)$ is Fermi function of electrons in the lead and
$\gamma_2$ is written for large separation from the Fermi level.
%%%This rate defines time $\tau_{\rm s}=1/\Gts$ after which
%%%initially unbalanced spin up and down states reach equilibrium populations.
The first term dominates when $\Delta \gtrsim -T$ ($T$ is the
temperature) and describes the probability of emptying a given spin
state back into the lead, after which the loading from the lead can
start again. This term is linear in the tunnelling rate
$\Gamma_\sm{R}$ between the quantum dot and its lead. The second
contribution $\gamma_2$ to the spin flip transition rate in
\req{eq:Tts}, commonly referred to as co-tunnelling, describes the
mechanism similar to the Korringa relaxation
mechanism\cite{korringa} of localized magnetic moments in bulk
conductors due to their coupling to the spins of itinerant
electrons. Note that although this term  contains two additional
small parameters, $T/|\Delta|$ and $\Gamma_\sm{R}/|\Delta|$, for a
deep level, \ie $\Delta\ll -T$, it vanishes only as $1/|\Delta|^2$,
compared to the exponential decay, $\exp(-|\Delta|/T)$,
of the first term.
We use \req{eq:Tts} to analyze experimental data
of Ref.\cite{shaji07} and obtain an almost perfect
agreement between the measured current and the current exclusively due to
extrinsic relaxation processes.

Previously, the influence of leads on processes in dots has been
addressed in connection to different phenomena, such as the Kondo
spin relaxation \cite{kaminski00},
co-tunneling~\cite{AverinNazarov92} transport and spin relaxation in
a single
dot~\cite{engel01,engel02,fujisawa02,sasaki95,dong06} and
nuclear spin relaxation in the Coulomb blockade regime
\cite{lyanda-geller02}.

The spin blockade occurs because transport through a double dot,
Fig.~\ref{fig:model}(a), can only flow via singlet states and is
blocked if electron entering the right dot from the right lead forms
a triplet with the spin on the left dot.\cite{fransson06,hanson07}
The relevant energy states corresponding to this regime are shown in
Fig.~\ref{fig:model}(b), where $(n_L,n_R)_{\rm S,T}$ denotes a
two-electron state with $n_{L(R)}$ electrons on the left (right) dot
in singlet (S) or triplet (T) configuration. States $(2,0)_{\rm
S,T}$ are separated by level spacing in the left dot, whereas
smallness of interdot exchange leaves states $(1,1)_{\rm S,T}$
nearly degenerate. Relaxation of the spin lifts the spin
blockade.\cite{tarucha06,inarrea07,koppens07} After time $\tau_{\rm
s}$ triplet state $(1,1)_{\rm T}$ can relax by spin flip into
singlet $(1,1)_{\rm S}$,
and thus the current $I\sim e/\tau_{\rm s}$ %in the blockade regime
can be used  to experimentally determine
the lifetime of an electron spin in a device, {\it c.f.} \cite{johnson05a}.
The importance of the spin relaxation due to electron exchange
with the leads is indicated by significant currents
on the boundaries of the spin blockade region that
has been observed in both GaAs\cite{johnson05}
and silicon\cite{shaji07,liu08} lateral double dots. % structures.
Moreover, this mechanism may dominate spin relaxation %be the most efficient one
in silicon quantum dots,
where intrinsic mechanisms are weak \cite{tahan02,zutic04,prada08}.
%We present quantitative analysis of spin-related dynamics due to lead effects. %these effects.

%\vspace*{0.4cm}

The full Hamiltonian for the double dot,
$\cH = {H}_{\rm d} + {H}_{\rm l}+ V$,
consists of interacting electrons in
the dots, ${H}_{\rm d}$, free electrons in the leads, ${H}_{\rm l}$,
and the tunnelling between the leads and the dots, $V$.
Here we do not specify the exact form of the Hamiltonian
$H_{\rm d}$ for strongly interacting electron states
in the double dot system, which in principle can
be written in terms of the creation, $d^\dag_{\alpha \,\sigma}$, and
annihilation, $d_{\alpha \,\sigma}$, operators in the left and
right dots ($\alpha=L,R$). We only assume that
the lowest eigenstates of ${H}_{\rm d}$ have the following
hierarchy of energies $E_{(2,0)_{\rm S}}<E_{(1,1)}<E_{(2,0)_{\rm T}}$,
as illustrated in Fig.~\ref{fig:model}(b).
The Hamiltonian of free electrons in the leads ${H}_{\rm l}$ is
written in terms of the creation, $c^\dag_{\alpha\,\vk\,\sigma}$, and
annihilation, $c_{\alpha\,\vk\,\sigma}$, operators of electrons in
lead $\alpha$ with momentum $\vk$, spin $\sigma$ and energy $\xi_{\alpha
\,\vk}$:
%\begin{equation}
$\displaystyle H_{\rm l}   = \sum_{\alpha=L,R} \sum_{\vk,\sigma}
\xi_{\alpha \,\vk} c^\dag_{\alpha\,\vk\,\sigma} c_{\alpha\,\vk\,\sigma} \,.$
%\label{eq:model}
%\end{equation}
The coupling between states in lead $\alpha$
to electron states in the
dot is represented by the tunnelling Hamiltonian $\hat V$,
with the tunnelling probabilities $ W_{\alpha \, k} $:
\begin{equation}
 V   = \sum_{\alpha=L,R} \sum_{\vk,\sigma} \left( W_{\alpha \,k}
d^\dag_{\alpha\,\sigma} c_{\alpha\, \vk \,\sigma} +
W^*_{\alpha \,k} c^\dag_{\alpha\, \vk \,\sigma} d_{\alpha \, \sigma} \right) \,.
\end{equation}

The relevant dot states that are involved in electron current are
single-particle states with an electron on the left dot,
$\ket{\uparrow 0} = d^\dag_{L\uparrow} \ket{0}$ ,
$\ket{\downarrow 0} = d^\dag_{L\downarrow} \ket{0}$,
and two-particle states with one electron on each dot,
that can be in either singlet,
$\ket{S} = \frac{1}{\sqrt{2}}
(d^\dag_{R\uparrow} d^\dag_{L\downarrow}
-d^\dag_{R\downarrow}d^\dag_{L\uparrow} )\ket{0}$,
or one of the triplet spin configurations,
$\ket{T_0} = \frac{1}{\sqrt{2}}
(d^\dag_{R\uparrow} d^\dag_{L\downarrow}
+d^\dag_{R\downarrow}d^\dag_{L\uparrow} )\ket{0}$,
$\ket{T_+} = d^\dag_{R\uparrow} d^\dag_{L\uparrow} \ket{0}$,
$\ket{T_-} = d^\dag_{R\downarrow} d^\dag_{L\downarrow} \ket{0} $.
The system can be in one of these states and
the corresponding probabilities satisfy the normalization:
$P_{\uparrow 0} + P_{\downarrow 0} + P_S + P_{T_0}  + P_{T_+} + P_{T_-} = 1$.
We omit the $(2,0)$ states, since $P_{(2,0)} \approx 0$ due to fast escape to left lead.
The rate equations for two-particle states in the dot are
\begin{widetext}
\begin{subequations}
\bea
\dot{P}_S &=& - (\gamma_1 +\frac{3}{2}\gamma_2 ) P_S
+\frac{1}{2} \bar \gamma_1 (P_{\uparrow0} + P_{\downarrow0})
+\frac{1}{2}\gamma_2 (P_{T_0} + P_{T_+} + P_{T_-}) - \Gamma P_S \,,\label{eq:PS} \\
\dot{P}_{T_0} &=& - (\gamma_1 + \frac{3}{2}\gamma_2 ) P_{T_0}
+\frac{1}{2}\bar \gamma_1 (P_{\uparrow0} + P_{\downarrow0})
+\frac{1}{2}\gamma_2 (P_{S} + P_{T_+} + P_{T_-}) \,, \\
\dot{P}_{T_+} &=& - (\gamma_1 +\gamma_2 ) P_{T_+}
+\bar \gamma_1 P_{\uparrow0} + \frac{1}{2} \gamma_2 (P_{S} + P_{T_0}) \,, \\
\dot{P}_{T_-} &=& - (\gamma_1 +\gamma_2 ) P_{T_-}
+\bar \gamma_1 P_{\downarrow0} + \frac{1}{2} \gamma_2 (P_{S} + P_{T_0}) \,.
\eea
\label{eq:rateeq}
\end{subequations}
\end{widetext}
For instance, \req{eq:PS} describes the change of the singlet
state population.
$P_S$ is reduced by the transitions to states
$\ket{\uparrow 0},\ket{\downarrow 0}$ (with rates $\gamma_1/2$),
but is increased by the reverse transitions (with rates $\bar\gamma_1/2$).
It also couples to all triplet states, $\ket{T_0}$ and $\ket{T_\pm}$,
with a single rate, $\gamma_2/2$.
Additionally, we include the possibility of transition from
singlet $\ket{S}=(1,1)_S$ to $(2,0)_{\rm S}$ between the two dots,
with rate $\Gamma$, by introducing $-\Gamma P_S$ term.
The remaining equations in system (\ref{eq:rateeq}) for triplet states
have similar structure, but without the interdot transitions.
%do not have terms for transition to $(2,0)$ states.

%%%%%%%%%%%%%%%%%%%%%%%%%%%%%%%%%%%%%%%%%%%%%%%%%%%%%%%%%%%%%%%%%%%%%%%%
\begin{figure}[t]
\includegraphics[width=0.85\linewidth]{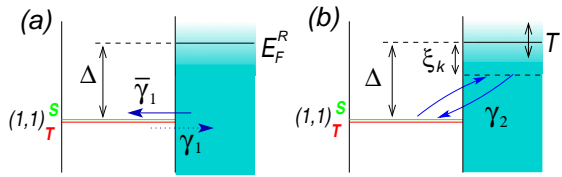}
\caption{ \label{fig:gamma12}
    (Color online) Electron exchange with the lead in lowest tunnelling orders.
    (a) In the first order process, a single electron jumps on or off the dot, with
    electron energy in the lead equal to $E_{(1,1)}$.
    The de-tuning from the Fermi level is $\Delta = E_{(1,1)} - E_F^R$.
    (b) The second order process represents
    double electron tunnelling through a virtual state.
    Energy of participating electrons in the lead are close to the lead Fermi energy
    and can be far from the dot energy level.
}
\end{figure}
%\noindent
%\hrulefill
%\end{wrapfigure}
%%%%%%%%%%%%%%%%%%%%%%%%%%%%%%%%%%%%%%%%%%%%%%%%%%%%%%%%%%%%%%%%%%%%%%%%

The transition rates in Eqs.(\ref{eq:rateeq}) are given by
the lowest two orders in tunnelling,
Fig.~\ref{fig:gamma12}, and describe
(a) electron hopping on~($\bar\gamma_1$) and off~($\gamma_1$) the dot
\be
\bar\gamma_1 = \Gamma_\sm{R} f(\Delta) \;,\quad \gamma_1 = \Gamma_\sm{R}
\left[1-f(\Delta)\right];
\label{eq:gamma1}
\ee
and (b) double exchange of electrons between the lead and the dot~($\gamma_2$),
via virtual states $\ket{\uparrow 0} ,\ket{\downarrow 0}$,
resulting in the creation of an electron-hole pair in the lead,
\be
\gamma_2 =  \frac{2\pi}{\hbar} \sum_\vk \cN_F |W_{R,k}|^4 \left|
\frac{1}{\Delta -\xi_{k} +i0} \right|^2 f(\xi_k)[1-f(\xi_k)] \,.
\nonumber
\ee
Here $f(\xi)=1/(1+\exp(\xi/T))$ is the
Fermi function in the right lead, and
$\Gamma_\sm{R} = (2\pi/\hbar) \, \cN_F \left| W_{R,k} \right|^2 $,
where $\cN_F$ is the density of states in the lead near the Fermi level
($\Delta,\xi \ll E_F$).
The expression for $\gamma_2$ formally diverges at $\xi_k = \Delta$, but is
applicable for $|\Delta|$ large compared with temperature,
when the contribution from $\xi_k \approx \Delta$
is exponentially suppressed %\cite{note}:
%%%%%%%%%%%%%%%%%%%%%%%%%%%%%%%%%
\footnote{The estimate in \req{eq:gamma2} becomes invalid
for $|\Delta| \lesssim T$. However the total spin relaxation rate is a sum of
$\gamma_1$ and $\gamma_2$, see \req{eq:spinrelax} later,
and $\gamma_1$ dominates the relaxation
in the range $\Delta \gtrsim -T \ln(T/T_\sm{R})$ with $T \gg T_\sm{R}$,
so $\gamma_2$ is to be omitted there anyway.
We write the rate equations assuming that these two rates
coexist in entire range of $\Delta$.
In the end, however, the contribution of $\gamma_2$
should be taken into account when $\Delta$ satisfies
the applicability condition of \req{eq:gamma2}.}:
%%%%%%%%%%%%%%%%%%%%%%%%%%%%%%%%%
\be
\gamma_2 \approx \Gamma_\sm{R} \frac{T \; T_\sm{R} }{\Delta^2} ,\quad
T_\sm{R} = \frac{\hbar \Gamma_\sm{R}}{2\pi}, \quad
|\Delta| \gtrsim T.
\label{eq:gamma2}
\ee
To conclude discussion of Eqs.~(\ref{eq:rateeq})-(\ref{eq:gamma2}),
we note that they were obtained
from general transition rate equations
for diagonal elements of the full density matrix.
Each such element corresponds to an eigenstate of the
full lead-dot system,
e.g. $\ket{i} = \ket{e_i} \times \ket{{\rm dot}_i}$ with energy $\epsilon_i$.
Transition rates between these states are,
\be
\Gamma^{fi} = \frac{2\pi}{\hbar} \delta(\epsilon_f - \epsilon_i)
\left| V_{fi} + \sum_m \frac{V_{fm} V_{mi}}{\epsilon_i-\epsilon_m +i0} +\dots \right|^2
\,.\nonumber
\ee
Since the environment relaxes much faster than the dot we
take the trace over the electronic configurations $\{e_i,e_f\}$
in the leads, and define
$\gamma_{fi} = {\rm Tr}_{e_i,e_f} \,
\Gamma^{fi} \, \rho^\sm{0}_{e_f} \rho^\sm{0}_{e_i}$
as the rate of transition between the \emph{dot} states
$\ket{{\rm dot}_i}$ and $\ket{{\rm dot}_f}$.
Here $\rho^\sm{0}_{e}$ is the equilibrium
density matrix for non-interacting electrons in the leads.

We assume that the lead-dot tunnelling rates $\Gamma_\sm{R,L}$
%between the right(left) lead and dot
are larger than the interdot rate $\Gamma$.
Indeed, the tunnelling between the
dots is accompanied by the emission of a phonon (or another
excitation) that carries away
the energy difference between $E_{(1,1)_{\rm S}}$
and $E_{(2,0)_{\rm S}}$. Such coupling of electron states to phonon
modes results in additional smallness of  the rate $\Gamma$, which
is determined by both the overlap between electron states in the two
dots and the matrix elements of electron-phonon coupling.
However, a microscopic derivation of $\Gamma$ is beyond the scope of this paper.
%in terms of microscopic characteristics of the system.

In the absence of magnetic fields,
states that differ only by spin projections, are degenerate,
and we introduce
$P_0 =  P_{\uparrow0} = P_{\downarrow0}$ and $P_{T_1} = P_{T_+} = P_{T_-}$.
We use the normalization to remove redundant $P_0$, and obtain
a system of equations for only three variables,
$P_{T_0}, P_{T_1}$ and $P_S$.

In the limit of negligible tunnelling to the left dot ($\Gamma\to0$),
this system of equations can be diagonalized:
%\be
%\dot P_{\eta}(t)+\Gamma_\eta P_\eta(t)=J_\eta, \eta=1,2,3:
%\ee
\bea
\dot P_{\eta}(t)+\Gamma_\eta P_\eta(t)=J_\eta \;,\quad \eta=1,2,3:
\qquad\quad
\label{eq:eigenmodes} \\ \nonumber
\begin{array}{ll@{\quad}l}
P_1 = P_{T_0} - P_{T_1}, &              \Gamma_1=\Gts, & J_1=0; \\
P_2 = 3 P_S - ( P_{T_0} + 2 P_{T_1}), & \Gamma_2=\Gts, & J_2=0; \\
P_3 = P_S + ( P_{T_0} + 2 P_{T_1} ), &  \Gamma_3 =\Gamma_{\rm c}, & J_3=2\bar\gamma_1.
\end{array}
\eea
The first two eigenmodes describe dynamics of
spin in the double dot system, with the
spin flip rate $\Gts$.
The last eigenmode is for %dynamics of %corresponds to the dynamics of
the total occupation of the right dot by one electron,
with characteristic charge relaxation rate $\Gamma_{\rm c}$.
The two rates are,
\be
\Gts= \gamma_1  + 2\gamma_2  \;, \quad
\Gamma_{\rm c}=\Gamma_\sm{R}[1+f(\Delta)].
\label{eq:spinrelax}
\ee

Here we remark on spin relaxation in a single dot.
The relevant states are empty ($P_0$)
and singly occupied with spin up/down ($P_{\uparrow/\downarrow}$).
One finds that Eqs.~(\ref{eq:rateeq}) for these states are modified,
but the dynamics of %eigenmode equations and the relaxation rates
$(P_\uparrow-P_\downarrow)$ and $(P_\uparrow+P_\downarrow)$
is given by \req{eq:eigenmodes} with $\eta=1$ and $\eta=3$ respectively,
with unchanged $\Gts$, $\Gamma_{\rm c}$ and the source terms.
%in \req{eq:spinrelax}.

\begin{figure}[t]
%\centerline{\includegraphics[width=0.9\linewidth]{spin_block_current.eps}}
\centerline{\includegraphics[width=0.9\linewidth]{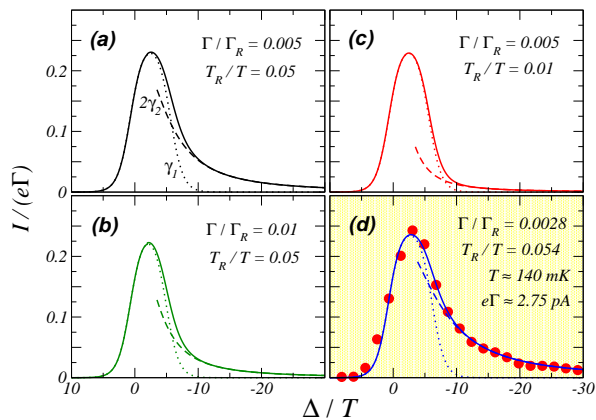}}
\caption{\label{fig:block} (Color online)
    Current through a double dot system in the spin blockaded regime.
    The dotted (dashed) lines show asymptotes due to first (second)
    order processes that dominate peak(valley) and are given by \req{eq:curr}
    with $\Gamma_{\rm s}$ replaced by
    $\gamma_1$ %, \req{eq:gamma1}
    ($2\gamma_2$). %(, \req{eq:gamma2}).
    Panel (d) shows a fit to measured current (circles)
    along a line-cut of the spin blockade peak
    reported in \cite{shaji07}.
}
\end{figure}

The current through the right dot is defined as the rate of charge escape
from the singlet state into the left dot,
$ %\be
I(t) = e \Gamma P_S(t) \,,
$ %\ee
and the stationary solution of rate equations, \reqs{eq:rateeq},
gives the steady current
\be
I = e \frac{f(\Delta)}{1+f(\Delta)}
\frac{2 \Gamma \Gts }{4\Gts + \Gamma ( 3   + \Gts/\Gamma_{\rm c}) }
 \,.
\label{eq:curr}
\ee
This  equation describes the magnitude of current through a lateral double
quantum dot in the regime of spin blockade. We plot typical current profiles
in Fig.~\ref{fig:block} as a function of de-tuning $\Delta$.
In experiments $\Delta$ is controlled by gate voltages and
current $I$ is mapped as a function of these voltages\cite{johnson05,shaji07}.

Equation (\ref{eq:curr}) has particularly simple form in two
important limits,
\begin{subequations}
\bea
\Gts \gg \Gamma \;: \;&& I=\onehalf e\Gamma \frac{f(\Delta)}{1+f(\Delta)}, \quad \mbox{(peak)} \,,
\label{eq:leftslope} \\
\Gts \ll \Gamma \;: \;&& I=\onethird e\Gamma_{\rm s} , \quad \mbox{(valley, $f(\Delta)\approx 1$)} \,.
\label{eq:rightslope}
\eea
\end{subequations}
The first limit describes the left slope and the peak of the
current, Fig.~\ref{fig:block}.
As a consequence, the dimensionless current $I/e\Gamma$
 is nearly independent of system parameters.
Its meaning is straightforward: the
right dot is loaded by one electron with probability
$2f(\Delta)/(1+f(\Delta))$, as follows from
Eqs.(\ref{eq:eigenmodes}) and (\ref{eq:spinrelax}),
and the singlet state is $1/4$ of this probability, as all two-particle states
are equally populated when singlet-triplet relaxation is fast.
The current is proportional to the escape rate $\Gamma$.
%of $(1,1)_{\rm S} \to (2,0)_{\rm S}$.

The second limit, \req{eq:rightslope}, describes the spin blockade region.
Depending on parameters it can exhibit distinct
non-exponentially decaying tail for $|\Delta|/T \gg 1$, see Fig.~\ref{fig:block},
that should be easily observable.
The factor $1/3$ comes from the probability of
the system to be in one of the triplet states.
Indeed, according to the stationary solution of Eqs.(\ref{eq:rateeq})
in the limit $|\Delta| \gg T$, the right dot is definitely occupied.
The probabilities of finding the system in the singlet and triplet states
are determined by the ratio of $\Gamma$ and $\Gts$,
\be
\left( \begin{array}{c}
P_S \\ P_{T_0} \\ P_{T_1}
\end{array}
\right)
= \frac{1}{ 4 \Gts + 3 \Gamma}
\left( \begin{array}{c}
\Gts \\ \Gts + \Gamma  \\ \Gts +
\Gamma
\end{array} \right) \,.\nonumber
\ee
For $\Gts \gg \Gamma$ the equilibrium is $P_S = P_{T_{0,1}} = 1/4$,
as we discussed.
On the other hand, if $\Gamma \gg \Gts$ the singlet state is
almost empty ($P_S \sim \Gts/3\Gamma$) since it takes
time $1/\Gts$ to populate this state from one of the triplet states
while it quickly empties into the left dot. The triplet states are
all equally populated, each with probability $\approx 1/3$.

%%%%%%%%%%%%%%%%%%%%%%%%%%%%%%%%%%%%%%%%%%%%%%%%%%%%%%%%%%%%%%%%%%%%%%%%
%%%%%%%%%%%%%%%%%%%%%%%%%%%%%%%%%%%%%%%%%%%%%%%%%%%%%%%%%%%%%%%%%%%%%%%%

Finally, we note that the second conduction peak observed in \cite{johnson05,shaji07}
is explained by complementary hole transport in the cycle
$(2,1)\to (1,1)_{\rm (T\to)S} \to (2,0)_S \to (2,1)$. In this cycle,
the spin flip relaxation between $(1,1)_{\rm T,S}$ states
occurs due to electron exchange between the left dot and
the left lead via intermediate states $(2,1)$.

%\paragraph{Conclusions.}
In conclusion, we presented a model for spin relaxation due
to electron exchange between dots and leads.
We used it to construct a theory of current through a spin blockaded
double quantum dot, where spin flips result in transitions between
$(1,1)_{\rm S}$ and $(1,1)_{\rm T}$ states,
with rate $\Gts = \gamma_1 + 2\gamma_2$. %see \req{eq:Tts}.
%This spin relaxation mechanism is important in devices when a dot
%is connected to the leads.
To estimate the resulting relaxation times,
we neglect intrinsic relaxation rate $\gamma_{\rm intr}$
(which in principle can be incorporated in \req{eq:curr} by replacement
$\gamma_2\to\gamma_2+\gamma_{\rm intr}$),
and use this theory to extract relevant times from experiment
on silicon double dots\cite{shaji07}.
We obtain $\tau_{\rm s} \sim 0.1$ ns in the peak of the current
and $\tau_{\rm s} \sim 1 \,\mu$s in the valley, see Fig.~3(d).
These relaxation times are comparable and even shorter than
those due to intrinsic spin-orbit and hyperfine mechanisms.
%Also, we point out that the intrinsic relaxation
%rate $\gamma_{\rm intr}$ can be incorporated
%to our result, \req{eq:curr}, by simple replacement
%$\gamma_2\to\gamma_2+\gamma_{\rm intr}$.
The coupling between electron states in the dots and the leads has
to be taken into account for analysis of intrinsic spin relaxation
mechanisms from the spin blockade measurements.

\paragraph{Acknowledgements.}
We would like to thank I. Aleiner, S. Coppersmith, M. Eriksson, M. Friesen,
L. Glazman, R. Joynt, M. Prada, and C. Simmons
for important discussions. We are also grateful to the authors of
Ref.~\cite{shaji07} for providing us with raw experimental data
used in Fig. 3 (d).

%%%%%%%%%%%%%%%%%%%%%%%%%%%%%%%%%%%%%%%%%%%%%%%%%%%%%%%%%%%%%%%%%%%
%%%%%\bibliographystyle{apsrev}
%%%\bibliographystyle{prsty}
%%%%%%\bibliography{NL2DEG}
%%%\bibliography{Meso}
%%%%%%%%%%%%%%%%%%%%%%%%%%%%%%%%%%%%%%%%%%%%%%%%%%%%%%%%%%%%%%%%%%%

\end{document}